\documentclass[preprint,12pt]{elsarticle}
\usepackage{amsmath,amssymb,amsfonts}
\usepackage{natbib}
\usepackage{algorithmic}
\usepackage{graphicx}
\usepackage[colorlinks=true, citecolor=blue]{hyperref}
\usepackage{textcomp}
\usepackage{enumitem}
\usepackage{caption}
\usepackage{hyperref}

\journal{NIM-A}

\date{\today}
\begin{document}

\begin{frontmatter}

\title{Quantitative Determination of Spatial Resolution and Linearity of Position-Sensitive LG-SiPMs at Sub-Millimeter Scale via Ricean Distribution Fitting}


\author[1]{Aramis Raiola\corref{cor1}}
\ead{aramis.raiola@unige.ch}
\author[4]{Fabio Acerbi}
\author[1]{Cyril Alispach} 
\author[2,3]{Hossein Arabi}
\author[1]{Domenico della Volpe}
\author[4]{Alberto Gola}
\author[2,3]{Habib Zaidi}

\affiliation[1]{
    organization={Départment de Physique Nucléaire et Corpusculaire (DPNC), Université de Genève}, 
    adressline={quai Ernest-Ansermet 24},
    postcode={1211}, 
    city={Genève 4}, 
    country={Switzerland}}
    
\affiliation[2]{
    organization={Département de Radiologie et Informatique Médicale, Université de Genève},
    adressline={rue Gabrielle-Perret-Gentil 4},
    postcode={1211},
    city={Genève 14}, 
    country={Switzerland}}
    
\affiliation[3]{
    organization={Service de Médecine Nucléaire et Imagerie Moléculaire, Hôpitaux Universitaires Genève (HUG)},
    adressline={rue Gabrielle Perret-Gentil 4}, 
    postcode={1205},
    city={Genève}, 
    country={Switzerland}}

\affiliation[4]{
    organization={Center for Sensor and Devices (SD), Fondazione Bruno Kessler (FBK)},
    adressline={Via Santa Croce 77},
    postcode={38122},
    city={Trento}, 
    country={Italy}}
\cortext[cor1]{Corresponding author}

%% Abstract
\begin{abstract}
Position-sensitive SiPMs are useful in all light detection applications requiring a small number of readout channels while preserving the information about the incoming light’s interaction position. Focusing on a 2x2 array of LG-SiPMs covering an area of $\sim 15.5\times15.5~\rm{mm}$ with just 6 readout channels, we proposed a quantitative method to evaluate image reconstruction performance. The method is based on a statistical approach to assess the device's precision (spatial resolution) and accuracy (linearity) in reconstructing the light spot center of gravity. This evaluation is achieved through a Rice probability distribution function fitting. We obtained an average sensor spatial resolution's best value of $81\pm3~\rm{\mu m}$ (standard deviation), which is achieved by reconstructing each position with the amplitude of the channels' output signals. The corresponding accuracy is of $231\pm4~\rm{\mu m}$.
\end{abstract}



%\item LG-SiPMs achieve sub-mm resolution over large areas with few channels.
%\item Iterative Rice fitting can quantify resolution and linearity across the device.
%\item LG-SiPMs suit scintillator optical readout in HEP and Nuclear Medicine when position sensitivity is required.
%% Keywords
\begin{keyword}
Light sensors\sep Silicon Photomultipliers \sep Position sensitive detectors \sep Nuclear medicine \sep High Energy Physics \sep Spatial resolution \sep Linearity correction



\end{keyword}

\end{frontmatter}



\section{Introduction}
The demand for sub-millimeter precision position-sensitive (PS) light sensors is a hot topic in various fields of modern experimental physics as well as for several industrial and/or medical applications~\cite{Schaefer1998, Del_Guerra2006, Acerbi2024, Zhang2023, Du2015, Schulz2013, surgeosight, McClish2010, Arabi_2024}. Given an incoming light beam, PS light detectors are devices designed to determine the light spot's center of gravity (CoG) position on the sensor's surface. Previous works, such as~\cite{Del_Guerra2006, Schaefer1998}, have already discussed the interest in these detectors in medical applications.
% One of PS light detectors' most promising future uses is compact gamma-ray cameras, or generally imaging scanners in nuclear medicine. 
By detecting the exact position of light flashes emitted by scintillator crystals, PS detectors effectively map the hit positions of an incident gamma-ray~\cite{Schulz2013,Du2015, Del_Guerra2006}. This allows for precise metabolism mapping in specific tissues within a living body previously tagged by radioactive isotopes.
\par In the context of scintillator crystal readout, very sensitive detectors, like Silicon Photomultipliers (SiPM) are often used. They have the advantage of high photodetection efficiency in the visible spectrum, compactness, insensitivity to magnetic fields, low bias voltages, and good time resolution. A particular type of PS SiPM device with position sensitivity is the so-called Linearly-graded (LG) SiPM technology patented by FBK~\cite{FBK-LG-SiPM, Gola2013ANA, Acerbi2024}.
\par In this work, we present a study on an array of 2x2 large-area LG-SiPMs~\cite{Acerbi2024}, where we developed a statistical method based on multiple Rice function fitting to determine precisely both the intrinsic resolution of the device and its linearity. With the latter term, we define the systematic shift between the physical position of the light source and the center of the reconstructed spot on the detector surface. A detector with poor reconstruction linearity will produce a deformed image. Thus, a preliminary assessment of the light sensor's linearity could be useful in developing a correction algorithm through statistical analysis or, more recently, by employing machine learning techniques and neural networks.
\section{Position-Sensitive LG-SiPMs}
\subsection{Position Sensitive Silicon Photomultiplier}
\par Among high-sensitive photodetectors, Silicon photomultipliers (SiPMs) are single-photon sensitive detectors that offer various advantages including high quantum efficiency, high gain, low bias voltages, and insensitivity to magnetic fields (which makes these devices suitable for applications in hybrid medical scanners, such as PET/MRI machines)~\cite{Gundacker_2020}. SiPMs are arrays of Single Photon Avalanche Diodes (SPADs) biased over the breakdown voltage (thus working in Geiger mode) and connected in parallel to have a common readout. The typical microcell size varies between few to tens of microns and, generally, it is possible to build SiPMs with active areas up to $10\times10~\rm{mm^2}$\cite{Gundacker_2020,Acerbi2024}. 
 Due to the Geiger mode operation of SiPMs, these devices are single photon sensitive. In each microcell (i.e. SPAD) the high electric field produces a relevant current pulse in response to a detected photon. The gain (i.e. number of carriers produced per each avalanche pulse) is in the order of $10^6$. 
\par There are two common approaches for realizing a SiPMs-based position-sensitive detector over large areas: building an array of several SiPMs or using position-sensitive SiPMs (PS-SiPMs). The first option requires numerous readout channels over large detection areas. Moreover, the resulting position resolution is limited by the size of the used SiPM chips. On the other hand, PS-SiPMs offer a sharper position resolution with fewer readout channels~\cite{Acerbi2024, Acerbi2024_ps}. 
\subsection{Linearly-Graded SiPMs}
\par Several PS-SiPMs approaches have been proposed in the last years, especially in the field of medical physics~\cite{Del_Guerra2006, Schaefer1998, FBK-LG-SiPM, Du2015, McClish2010, Schulz2013, Gola2020}. In the present work, we used the linearly graded SiPMs (LG-SiPM) technology, developed by FBK, both with NUV-HD SiPMs technology~\cite{FBK-NUV-SiPM, Acerbi2024} (with higher photodetection efficiency toward blue-ultraviolet light) and RGB-HD SiPMs~\cite{FBK-RGB-SiPM, Acerbi2024} (targeting visible light). These light sensors were mainly tested in the context of small-animal PET scanners~\cite{FBK-LG-SiPM, Du2015}, even though other applications for High-Energy physics were proposed~\cite{Gola2020}. LG-SiPMs working principle is based on a weighted avalanche charge division through two linearly-graded current dividers. At the detection of a photon, the charge of each microcell avalanche is divided into two paths through a current divider (one for vertical coordinate and one for horizontal coordinate).%, where the conductance is proportional to the local position of the firing SPAD. 

\begin{figure*}
\vspace{-0.3cm}
  \centering
  \begin{minipage}[!t]{0.49\textwidth}
    \centering
    \includegraphics[width=0.9\textwidth]{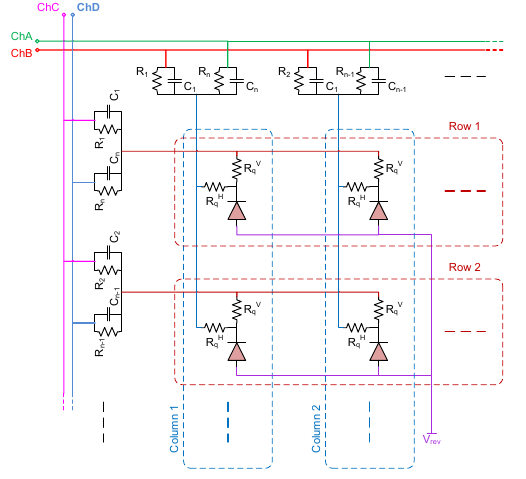}
    \caption{Scheme showing a portion of an LG-SiPMs microcells connected to the resistive/capacitive network, which allows for position encoding~\cite{Acerbi2024}.}
    \label{fig:LG-SiPM}
  \end{minipage}
   \hfill
    \begin{minipage}[!b]{0.49\textwidth}
    \centering
\includegraphics[width=0.9\textwidth]{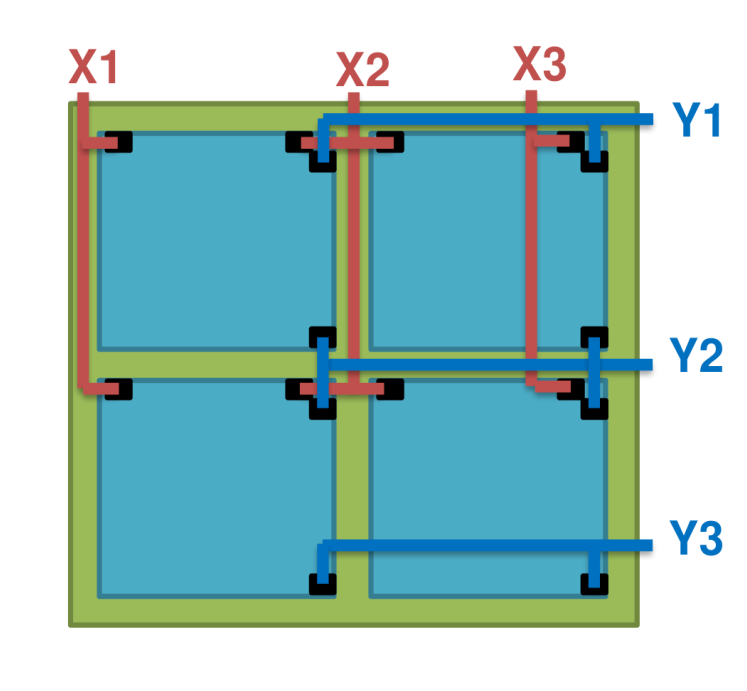}
    \caption{2x2 LG-SiPMs array with readout channels scheme (smart channel layout)~\cite{Acerbi2024}. The charge read out from these 6 channels is the input for equation~\ref{eq:2x2array}.}
    \label{fig:smart_c}
  \end{minipage}
\end{figure*}

The two split output currents are further divided by the current dividers connected to the two readout channels (e.g. ChA and ChB for the horizontal direction in figure~\ref{fig:LG-SiPM}). These current dividers allow the reconstruction of the triggered SPAD position. This is possible thanks to vary to produce a linearly increasing conductivity from the 0-th to the n-th node~\cite{FBK-LG-SiPM,Acerbi2024}. The amplitude and charge of the four output signals allow for determining the chip's light interaction location, which corresponds to the triggered SPAD position in the network.

If the top-left cell is fired in the horizontal direction, a large amount of charge will be collected from channel \(Ch_B\), while a smaller current flows through \(Ch_A\), whose branch resistance is higher. The situation is reversed when the outermost top-right SPAD is fired. More precisely, the difference between the output waveforms of \(Ch_A\) and \(Ch_B\) (both in terms of amplitude or total charge) is linearly varying depending on the horizontal position of the triggered SPAD. This principle also applies to the vertical direction.

The 2D position of the fired SPAD for a single LG-SiPM can be, therefore, reconstructed using the Anger camera CoG equations~\cite{AngerCamera,FBK-LG-SiPM, Acerbi2024}:

\begin{equation}
x_{rec}=\frac{l_x}{2}\frac{Q_A-Q_B}{Q_A+Q_B};~y_{rec}=\frac{l_y}{2}\frac{Q_C-Q_D}{Q_C+Q_D}
\label{eq:LG-SiPM}
\end{equation}

where \(x\) and \(y\) are the normalized horizontal and vertical positions, \(l_x\) and \(l_y\) are the dimensions of the SiPM chip in the respective directions, and \(Q\) represents the charge collected by each channel, which can be extracted from the signal waveform using different techniques. Thus, position reconstruction with LG-SiPMs can be achieved using just 4 channels for the entire device area (usually less than \(10 \times 10 \: \text{mm}^2\)), with a precision that virtually corresponds to one microcell diameter~\cite{FBK-LG-SiPM,Acerbi2024}.

In the case of an extended incoming beam (when multiple cells are simultaneously fired within a single SiPM), the current output from the readout channels can still be processed through equation~\ref{eq:LG-SiPM}. In this scenario, the reconstructed position corresponds to the CoG position of the light beam, i.e., the geometrical center of the incoming light (for a uniform beam). The reconstruction resolution is typically a few hundred microns due to several electronic aberrations that perturb the position reconstruction and do not allow it to reach single-microcell resolution. 

\begin{figure*}[h]
    \centering
    \includegraphics[width=\textwidth]{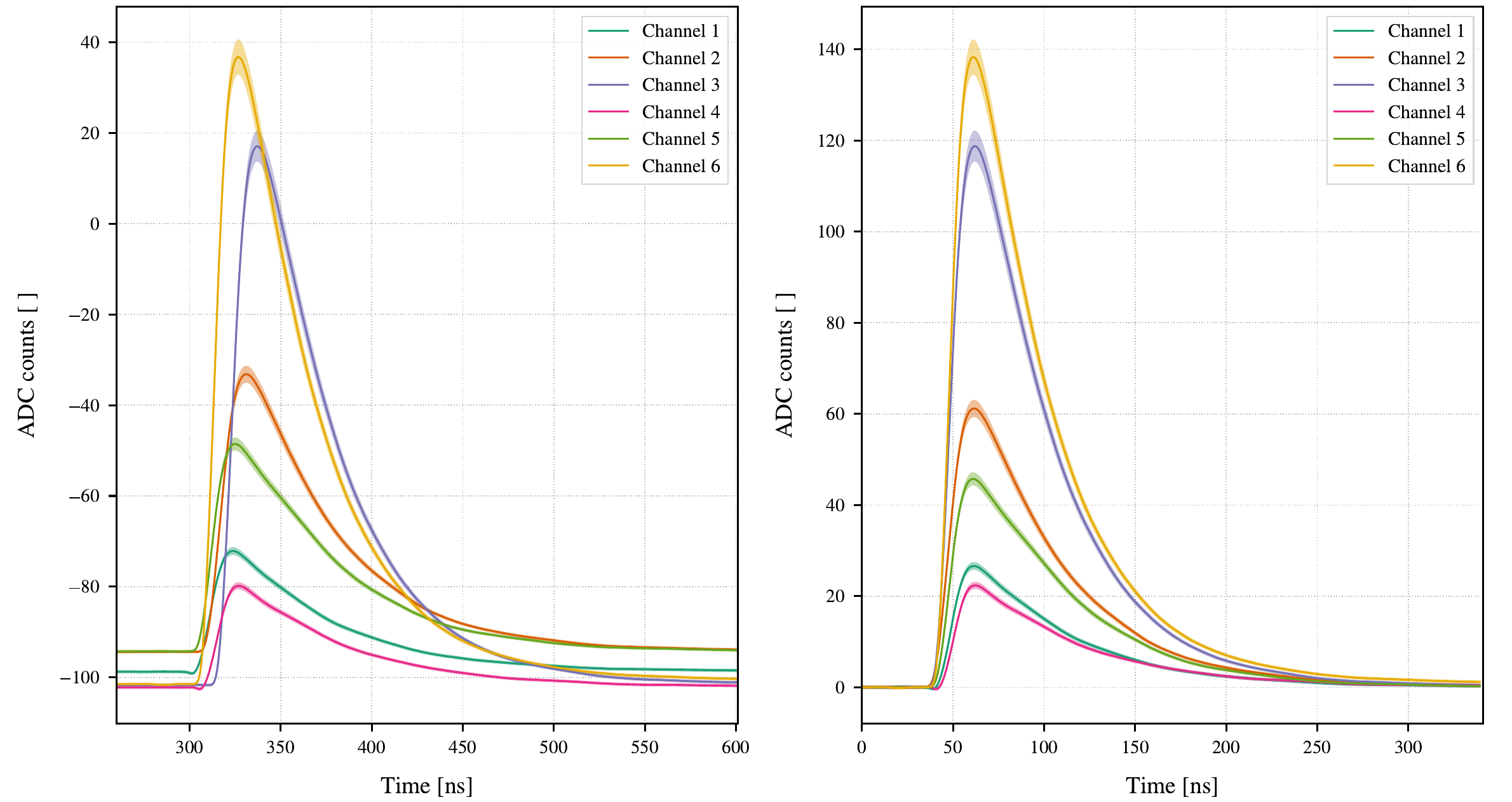}
    \caption{Average LG-SiPMs array response waveforms for each channel for the single motor position (29.0 mm; 10.5 mm) out of the 10'000 positions recorded. On the left are shown the acquired waveforms for each channel and on the right the same waveforms after delay and baseline corrections}
    \label{fig:wave}
\end{figure*}

\subsection{2x2 array of LG-SiPMs}
LG-SiPMs can also be arranged in arrays of multiple devices to increase the detector's active surface and have a larger field of view (FOV). 
In this work, we used a 2x2 array of RGB-HD-based LG-SiPMs, i.e., based on a p-type substrate, having a photon detection efficiency (PDE) peaking in the green wavelength range. The tested device has a single-chip area of $7.75x7.75~\rm{mm^2}$~\cite{Du2018}, with $20~\rm{\mu m}$ SPAD pitch and a breakdown voltage of about $28~\rm{V}$. 

\par The array is an assembly of 4 SiPM chips connected with a "smart-channel" configuration ~\cite{Acerbi2024}. In this configuration, two adjacent chips' output is connected, forming a single middle channel (channels $x_2$ and $y_2$ in figure~\ref{fig:smart_c}). This layout allows to reduce further the number of readout channels~\cite{Gola2020, Acerbi2024}. 
For an \(n \times n\) array of LG-SiPMs, the number of channels scales as: \(n_{\text{channels}} = 2 \cdot (n + 1) \), compared to the \(n^2\) for arrays of standard analog SiPMs and the \(4 \cdot n\) trend for LG-SiPMs without the smart-channel configuration. 

However, this further reduction of output lines comes with the cost of a slight worsening of position resolution because of the increased parasitic \cite{Acerbi2024}. 
\par The position reconstruction for a 2x2 array of PS SiPMs can be calculated with the following variation of equation~\ref{eq:LG-SiPM}:
\begin{equation}
    x_{rec}=\frac{l_x}{2}\frac{x_3-x_1}{x_1+x_2+x_3};~y_{rec}=\frac{l_y}{2}\frac{y_3-y_1}{y_1+y_2+y_3}
    \label{eq:2x2array}
\end{equation}
where $x_i$ and $y_i$ ($i=1,2,3$) are the charge output of the readout channels (as displayed in figure~\ref{fig:LG-SiPM}) and $l_x$ and $l_y$ are the lateral sizes of the whole array. For more details about LG-SiPMs, cf.~\cite{FBK-LG-SiPM, Acerbi2024}.

\section{Experimental procedure and data reduction}
\subsection{Testing setup}
\begin{figure}[h]
    \centering
    \includegraphics[width=0.6\linewidth]{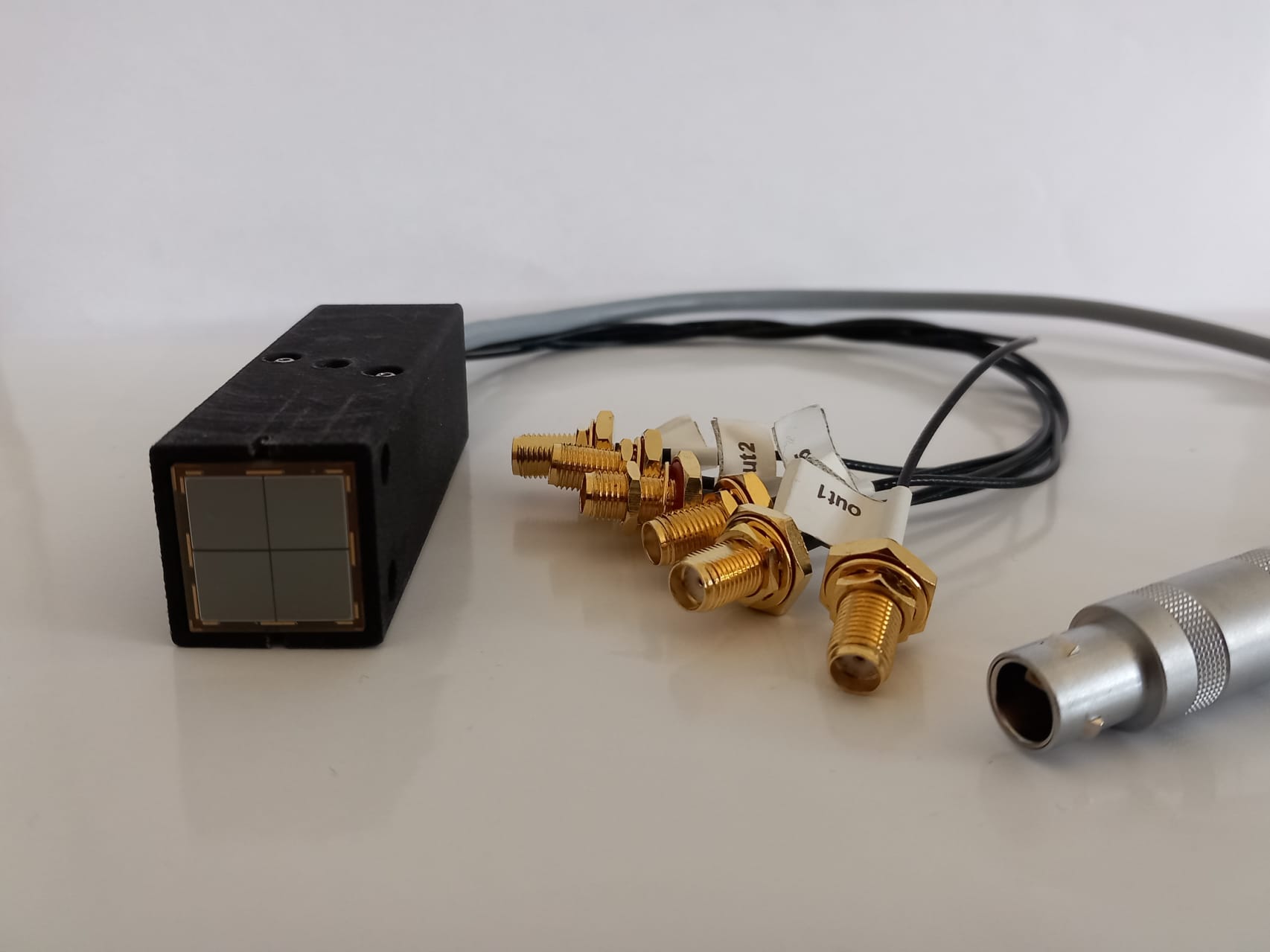}
    \caption{Picture of the tested LG-SiPM array module including a 2x2 array of LG-SiPMs and the 6-channels amplifier board, with 6 output signals (SMA connectors).}
    \label{fig:POSiCS-I}
\end{figure}

To assess the position resolution and linearity of the 2x2 RGB-HD LG-SiPMs array under test (Fig.~\ref{fig:POSiCS-I}), we scanned its active surface with a pulsed blue LED (central wavelength: 470 nm). A pulse generator (Aim-TTi TG5011) drives the light source by sending squared $4~\rm{V}$ amplitude pulses at a $1~\rm{kHz}$ rate and with a width of $20~\rm{ns}$. The LED light pulses were sent on the device through a multi-mode optical fiber (M92L02), whose exit was placed at $1~\rm{mm}$ from the SiPMs. It is possible to estimate from the fiber specifications that the light spot diameter that we expect on the detector's surface is about $2~\rm{mm}$.
\par The LG-SiPMs array was mounted on 2 Thorlabs linear stages (mts 50/m-z8) to control the device in the x and y directions on the plane parallel to the sensor's active surface. 

\par With this setup, we performed a 37x37 point scan of the SiPM array surface in a grid pattern, with a step of $0.5~\rm{mm}$. We illuminated the sensor with 10,000 pulses for each position, and an equal number of positions were reconstructed through the previously reported equation~\ref{eq:2x2array}. 
\par At each step, we recorded the analog signals from the six channels using a 12-bit ADC oscilloscope (Tektronix MSO58LP) with 8 channels and $1~\rm{GHz}$ bandwidth. The SiPM $V_{ov}$ was set at $5~\rm{V}$, and the tests were performed at room temperature $20^\circ~\rm{C}-25^\circ~\rm{C}$. 

\subsection{Charge-based event selection}
We performed offset correction and charge-based event selection to identify and remove weak signals from partially illuminated light spots, which can lead to shifted reconstructed positions.

\par First, waveform outputs were corrected for baseline offset and temporal alignment. The baseline was averaged between $0~\rm{ns}$ and $200~\rm{ns}$ and subtracted from all pulse charges. Temporal alignment accounted for horizontal peak shifts due to varying cable lengths (Fig.~\ref{fig:wave}).

\par A threshold selection was applied to the integral charge of the waveforms to exclude events that only partially detected the light spot, such as those near the edges or partially falling into the cross-shaped gap between the SiPMs. The number of reconstructed positions surviving this selection is 64.9\% of the original set, with an average of 6489 events per motor position.

\subsection{Position reconstruction}
The physical position reconstruction has two main goals: determining the scales $l_x$ and $l_y$ in equation~\ref{eq:2x2array} and aligning SiPM coordinates with linear stage coordinates. 

\par The first step addresses the limited range of normalized reconstructed positions, which do not span $[-1, 1]$ due to parasitic impedances and signal extraction non-idealities \cite{Acerbi2024}. Therefore, we determined two effective extended lengths $l_{x,\rm{eff}}$ and $l_{y,\rm{eff}}$ for reconstructed positions, ensuring accurate reconstruction resolution. The second step aligns reconstructed positions with the motor’s linear stage coordinates. The motor’s "home position" $(0~\rm{mm}, 0~\rm{mm})_{motor}$ does not coincide with the reconstructed center $(0~\rm{mm}, 0~\rm{mm})_{rec}$ due to misalignment between the device under test and the linear stage. Thus, correcting for this shift is crucial for reliably estimating position linearity. These parameters were determined by developing a minimization model that minimally reduces the difference between the coordinates registered by the linear stage and those computed from the LG-SiPM waveforms. 

\par Let us define the normalized amplitude vector $\Vec{x}\in\mathbb{R}^6$: 
\begin{align}
\Vec{x} &= \left(\frac{x_1}{x_1+x_2+x_3}, \frac{x_2}{x_1+x_2+x_3}, \frac{x_3}{x_1+x_2+x_3}, \right. \nonumber \\
&\quad \left. \frac{y_1}{y_1+y_2+y_3}, \frac{y_2}{y_1+y_2+y_3}, \frac{y_3}{y_1+y_2+y_3}\right)
\end{align}
These values correspond to the normalized charge output of the 6 channels encoding the event's x- and y-positions. The following section provides more details on how to extract this charge. 
We then define the reconstructed positions vector $(x_{rec}, y_{rec})\in\mathbb{R}^2$ and link it the normalized charge output:

\begin{equation}
    \begin{pmatrix} x_{rec} \\ y_{rec} \end{pmatrix} \; =A\Vec{x}+ \Vec{b} \;
    \label{eq:linar_FBK}
\end{equation}

Where $\Vec{b}=(x_0,y_0)$ corresponds to the offset between the reconstruction coordinates, (for which the  $(0~\rm{mm},0~\rm{mm})_{rec}$ is the array's central position) and the linear stage coordinates. The matrix $A\in\mathbb{R}^{2\times6}$ is then defined as:

\begin{equation}
    A=\begin{pmatrix}
l_{\rm{x,eff}}/2 & 0 & -l_{\rm{x,eff}}/2  & 0 & 0 & 0 \\
0 & 0 & 0 & l_{\rm{y,eff}}/2  & 0 & -l_{\rm{y,eff}}/2  
\end{pmatrix}
\end{equation}

Where $l_{\rm{x,eff}}$ and $l_{\rm{y,eff}}$ are the effective lateral lengths that the minimization algorithm will determine. We introduced another parameter, $\varphi$, to complete the model. This rotation angle along the device surface was considered since, in our practical setup implementation, the linear micro-positioner stages and the detector have been aligned manually without precise references. A tilt of a few degrees between the two is possible and must be considered to minimize the difference between LED nominal positions and reconstructed light spots CoG. To align the two reference systems (motor and reconstructed), we applied to the reconstructed position vector a 2D rotation matrix:

\begin{equation}
    R(\varphi)=\begin{pmatrix}
\cos(\varphi) & -\sin(\varphi) \\
\sin(\varphi) & \cos(\varphi)
\end{pmatrix}
\end{equation}

linearly transforming equation \ref{eq:linar_FBK} into a tilt-dependent coordinates transformation: 

\begin{equation}
\begin{pmatrix} x'_{rec} \\ y'_{rec} \end{pmatrix} \;=R(\varphi)\begin{pmatrix} x_{rec} \\ y_{rec} \end{pmatrix} \;=R(\varphi)A\Vec{x}+R(\varphi)\Vec{b}
\end{equation}

The equations then represent the final model of position reconstruction to minimize:
\fontsize{8}{10}\selectfont
\begin{subequations}
\begin{equation}
x'_{rec}=\cos(\varphi)\left(\frac{l_{\rm{x,eff}}}{2}\frac{x_3-x_1} {x_1+x_2+x_3} +x_0\right) - \sin(\varphi) \left(\frac{l_{\rm{y,eff}}}{2}\frac{y_3-y_1}{y_1+y_2+y_3} +y_0\right)
\end{equation}   

\begin{equation}
y'_{rec}=\sin(\varphi)\left(\frac{l_{\rm{x,eff}}}{2}\frac{x_3-x_1} {x_1+x_2+x_3} +x_0 \right)+\cos(\varphi)\left( \frac{l_{\rm{y,eff}}}{2}\frac{y_3-y_1}{y_1+y_2+y_3} + y_0\right)
\end{equation}
\label{eq:POSICS-I_modified}
\end{subequations}
\normalsize
With this model, we can perform the minimization of the distance value $d$:

\begin{equation}
    d=\sqrt{(x'_{rec}-x_{true})^2+(y'_{rec}-y_{true})^2}
    \label{eq:new_d}
\end{equation}

representing all the distances between reconstructed points and the motor position given by the Thorlabs linear stage output. The minimization was implemented through the Scipy minimizer library applied on the free parameters: $l_{\rm{x,eff}}$, $l_{\rm{y,eff}}$, $x_0$, $y_0$, $\varphi$, with the Nelder-Mead method, suited for the minimization of a multi-variable scalar function. This method does not rely on gradient descent and efficiently minimizes non-smooth functions. However, since it does not follow the function gradient for minimization, it is challenging to directly estimate an inverse Hessian matrix for uncertainty determination. Thus, uncertainties were estimated by randomly sampling the reconstructed events and repeating the minimization multiple times. The estimated uncertainty corresponds to the standard deviation of the bootstrapped parameters distribution. The final parameters results will then provide a model to compare the reconstructed positions in $mm$ with the motor coordinates and estimate the PS-SiPM spatial resolution and linearity.
\subsection{Rice distribution fitting}
\begin{figure*}
\vspace{-0.3cm}
  \centering
  \begin{minipage}[!t]{0.49\textwidth}
    \includegraphics[width=\textwidth]{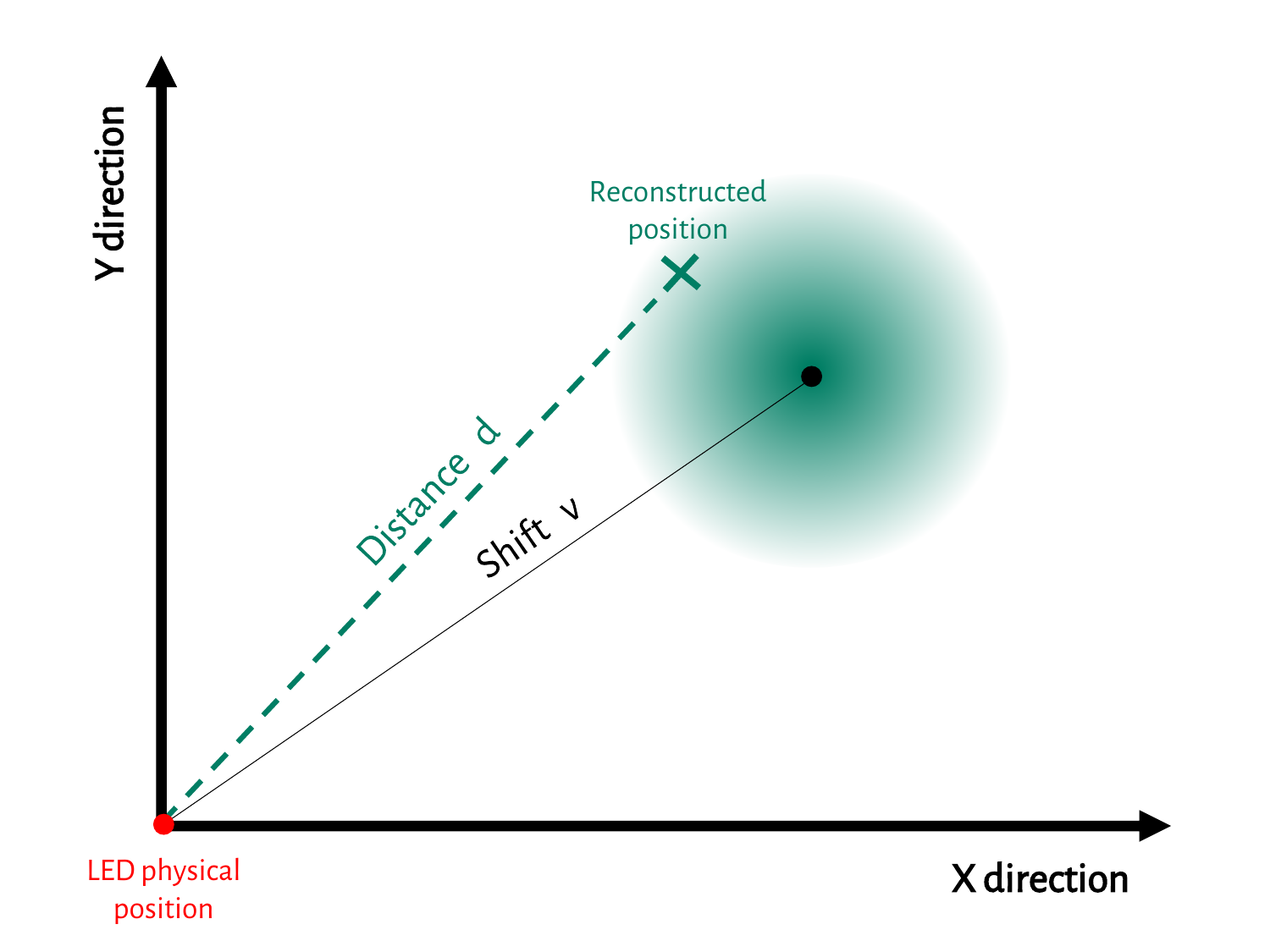}
    \caption{Scheme of a possible spread of the 10'000 reconstructed position (green halo) following a non-central 2D Gaussian distribution.}
    \label{fig:rice_schematics}
  \end{minipage}
   \hfill
    \begin{minipage}[!b]{0.49\textwidth}
    \includegraphics[width=\textwidth]{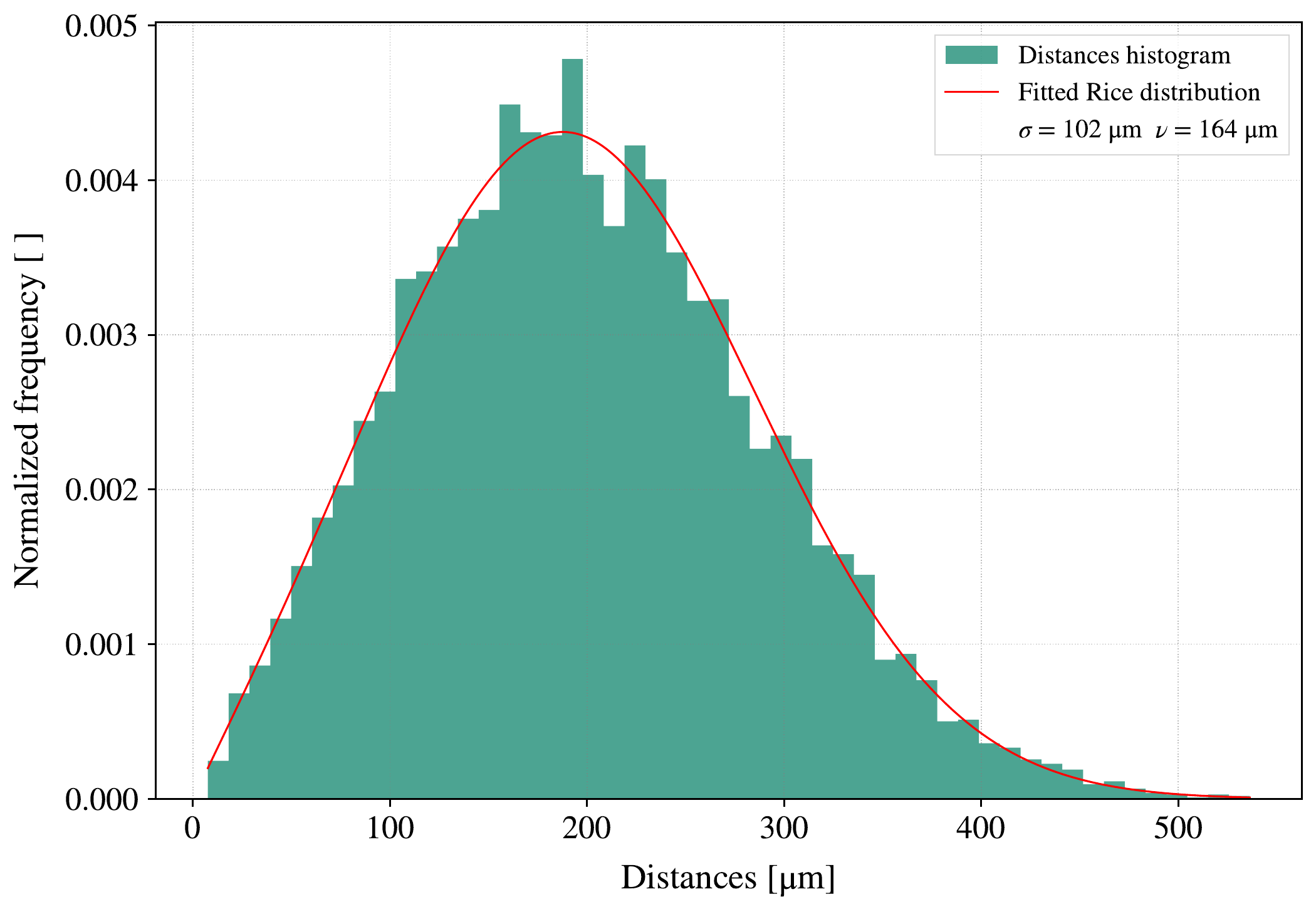}
    \caption{Rice fit on the distribution of distances $d$ from the true motor position. This sample contains the LED pulses for a single motor position.}
    \label{fig:ricean_fit}
  \end{minipage}
\end{figure*}

Once the minimization is carried out, we obtain a distribution of reconstructed points per each motor position (typically spread around a central value). The assumption done in this test is that the points are spread in space as a 2D Gaussian distribution, with the same standard deviation $\sigma$ both in the x and y directions. This is typically true towards the camera's center, while the points at the camera's border tend to show a certain degree of oblateness due to different image deformation effects.  
The mathematical distribution used to model the points spread is known as \textit{Rice distribution}. This probability density function (PDF) describes the dispersion of two Gaussian variables, $X$ and $Y$, sharing the same standard deviation. Given two Gaussian variables $X\sim \mathcal{N}(\mu_x,\sigma)$ and $Y\sim\mathcal{N}(\mu_y,\sigma)$, a Rician random variable is defined as:
\begin{equation}
    \mathcal{R}(\nu,\sigma)\sim\sqrt{X^2+Y^2}
\end{equation}
where $\nu$ represents the shift between the center of coordinates and the Gaussian peak: $\nu=\sqrt{\mu_x^2+\mu_y^2}$~\cite{rice} (cf. figure~\ref{fig:rice_schematics}). Computing with equation~\ref{eq:new_d} the distance $d$ between the reconstructed positions for the same motor position and the motor position given by the linear stage, we then suppose that they follow a Rice distribution:

\begin{equation}
    P(d|\nu,\sigma)=\frac{d}{\sigma^2}\exp\left(-\frac{d^2+\nu^2}{2\sigma^2}\right)I_0\left(\frac{\nu d}{\sigma^2}\right)
\end{equation}

where $I_0$ is the zeroth order modified Bessel function~\cite{rice}. By performing a Ricean fit on the $d$ distribution for each motor position, we can assess the shift $\nu$ (linearity parameter) and the standard deviation $\sigma$ (resolution) of the LG-SiPMs array at each position on its surface, as done in figure~\ref{fig:ricean_fit}. 

\section{Testing results}
A possible method to extract the SiPMs channels' amplitude output to reconstruct the incoming light central position is to sample the waveforms' amplitude from the 6 channels and input them in equation~\ref{eq:2x2array}. Then, the minimization can be performed by finding the 5 model's free parameters, following equation~\ref{eq:POSICS-I_modified}. 
\par The minimization was carried out to find the parameters presented in table~\ref{table:pars}.

The statistical uncertainties determined through bootstrapping methods were extremely small ($\sim1~\rm{\mu m}$) due to a large number of points constraining the minimization. Therefore, the presented uncertainties are the summation in quadrature of the fit (statistical) uncertainty and the systematic errors due to the motor positioning: the bidirectional repeatability (6~$\rm{\mu m}$) and backlash ($\sim1\rm{\mu m}$). 
\begin{table}[h!]
\renewcommand{\arraystretch}{1.2}  
\centering
\begin{tabular}{p{0.5\linewidth}c}
\hline\hline
Parameter           & Minimization result \\ \hline
Effective extended length $l_{x,\rm{eff}}$ & 21.049 $\pm$ 0.006 mm\\
Effective extended length $l_{y,\rm{eff}}$ & 21.112 $\pm$ 0.006 mm\\
Coordinates shift $x_0$ & 15.190 $\pm$ 0.006 mm\\
Coordinates shift $y_0$ & 23.074 $\pm$ 0.006 mm\\
Tilt angle $\varphi$ & 0.006$^\circ$ $\pm$ 0.001$^\circ$\\
\hline
\end{tabular}
\caption{Minimization results for the waveform amplitude reconstruction method.}
\label{table:pars}
\end{table}
The reconstructed image processed with the minimization parameters has a lateral size of $L_x=15.284~\rm{mm}$ and $L_y=15.020~\rm{mm}$. We define the size ratio $R_s$ as the ratio between the reconstructed average lateral size of the device and its physical lateral size $L_0$:
\begin{equation}
    R_s = \frac{1}{2}\frac{(L_x+L_y)}{L_0}=5.4\%
\end{equation}
which will be useful in computing errors on the spatial resolution and linearity of the detector. The fact that we obtain only a small fractional difference between the physical SiPMs array size and the reconstructed one makes the minimization results more reliable.
\begin{figure}[h!]
    \centering
    \includegraphics[width=0.7\linewidth]{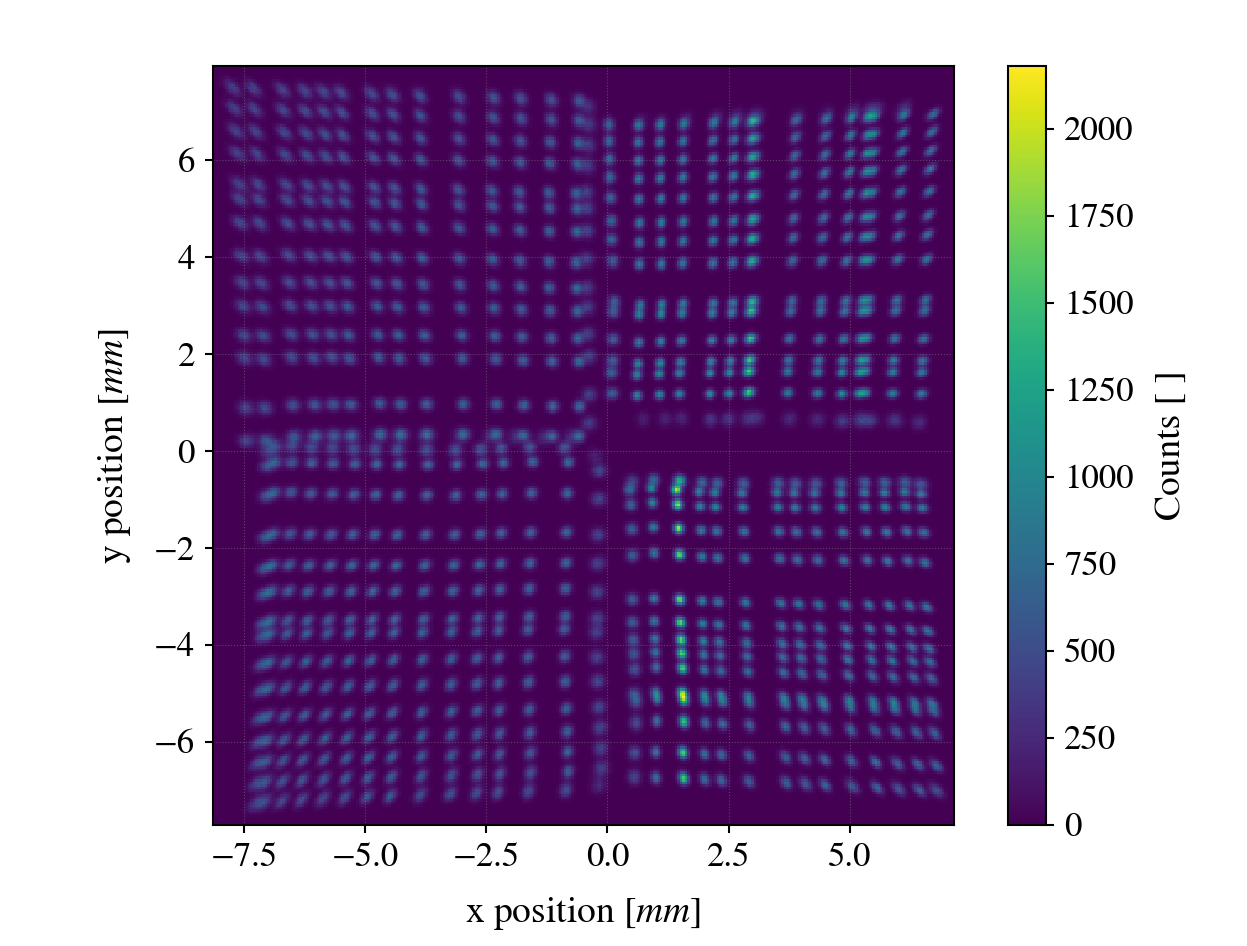}
    \caption{Reconstructed positions 2d histogram after minimization.}
    \label{fig:amplitude_method}
\end{figure}
With the minimization parameters, we compute the distance between the nominal and reconstructed positions with equation~\ref{eq:new_d} for each motor position. Subsequently, a histogram of these distances is created for each motor position. A Rice distribution is then fitted to each histogram to analyze the detector’s resolution (standard deviation $\sigma$) and linearity parameter (shift $\nu$) at different points of the detector's surface.
Additionally, the fit results on each point were averaged (weighted average with variance-defined weights) to give a proxy of the resolution on the whole camera surface:
    \begin{subequations}
    \begin{equation}
        \bar{\sigma}\pm\Delta\bar{\sigma} = \frac{\sum_{i,j}\sigma_{i,j}w_{i,j}^\sigma}{\sum_{i,j}w_{i,j}^\sigma} \pm \sqrt{\frac{1}{\sum_{i,j}w_{i,j}^\sigma}}
        \label{eq:sigma}
    \end{equation}
    
    \begin{equation}
        \bar{\nu}\pm\Delta\bar{\nu} = \frac{\sum_{i,j}\nu_{i,j}w_{i,j}^\nu}{\sum_{i,j}w_{i,j}^\nu} \pm \sqrt{\frac{1}{\sum_{i,j}w_{i,j}^\nu}}
        \label{eq:nu}
        \end{equation}
    \end{subequations}
Where $i,j$ are the rows and columns index of the motor positions grid, $w_{i,j}^\sigma~=~1/\Delta\sigma_{i,j}$ and $w_{i,j}^\nu~=~1/\Delta\nu_{i,j}$. The statistical uncertainty on the fit parameters was once again obtained through Bootstrap. 
\par The resolution errors at each motor position are calculated as the quadrature sum of the systematic error $\Delta\sigma_{syst}$ and the statistical error $\Delta\sigma_{stat}$. Statistical uncertainty corresponds to the fit error (from the covariance matrix), while the systematic error is defined as the product of the fitted $\sigma$ and the size ratio:
\begin{equation}
    \Delta\sigma = R_s \cdot \sigma \oplus \Delta\sigma_{fit}
\end{equation}
The systematic error accounts for deviations in the reconstructed image size compared to the physical array, which could under/over-estimate the 2D Gaussian standard deviation. This estimation also applies to the shift parameter, $\nu$, which depends on the size ratio and motor position. However, $\nu$ is further affected by the linear stage backlash (BL) and bidirectional repeatability (BR):
\begin{equation}
    \Delta\nu = \Delta\nu_{BL} \oplus \Delta\nu_{BR} \oplus R_s \cdot \nu \oplus \Delta\nu_{fit}
\end{equation}
\begin{figure*}
\vspace{-0.3cm}
  \centering
  \begin{minipage}[!t]{0.49\textwidth}
    \includegraphics[width=\textwidth]{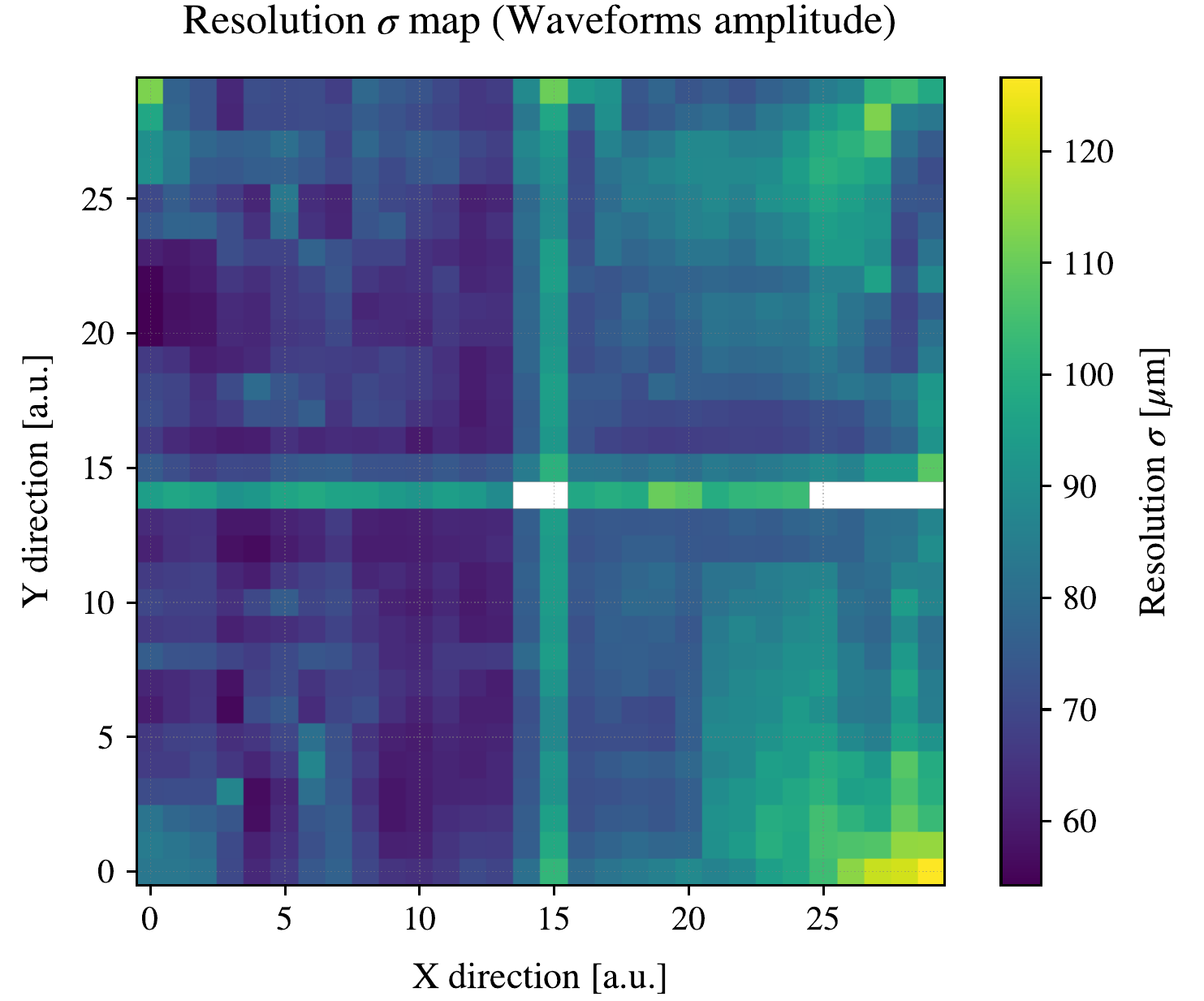}
    \caption{Resolution map of the SiPM surface scan using Rice fitting. Each bin represents a motor step.}
    \label{fig:res_amp}
  \end{minipage}
   \hfill
    \begin{minipage}[!b]{0.49\textwidth}
    \includegraphics[width=\textwidth]{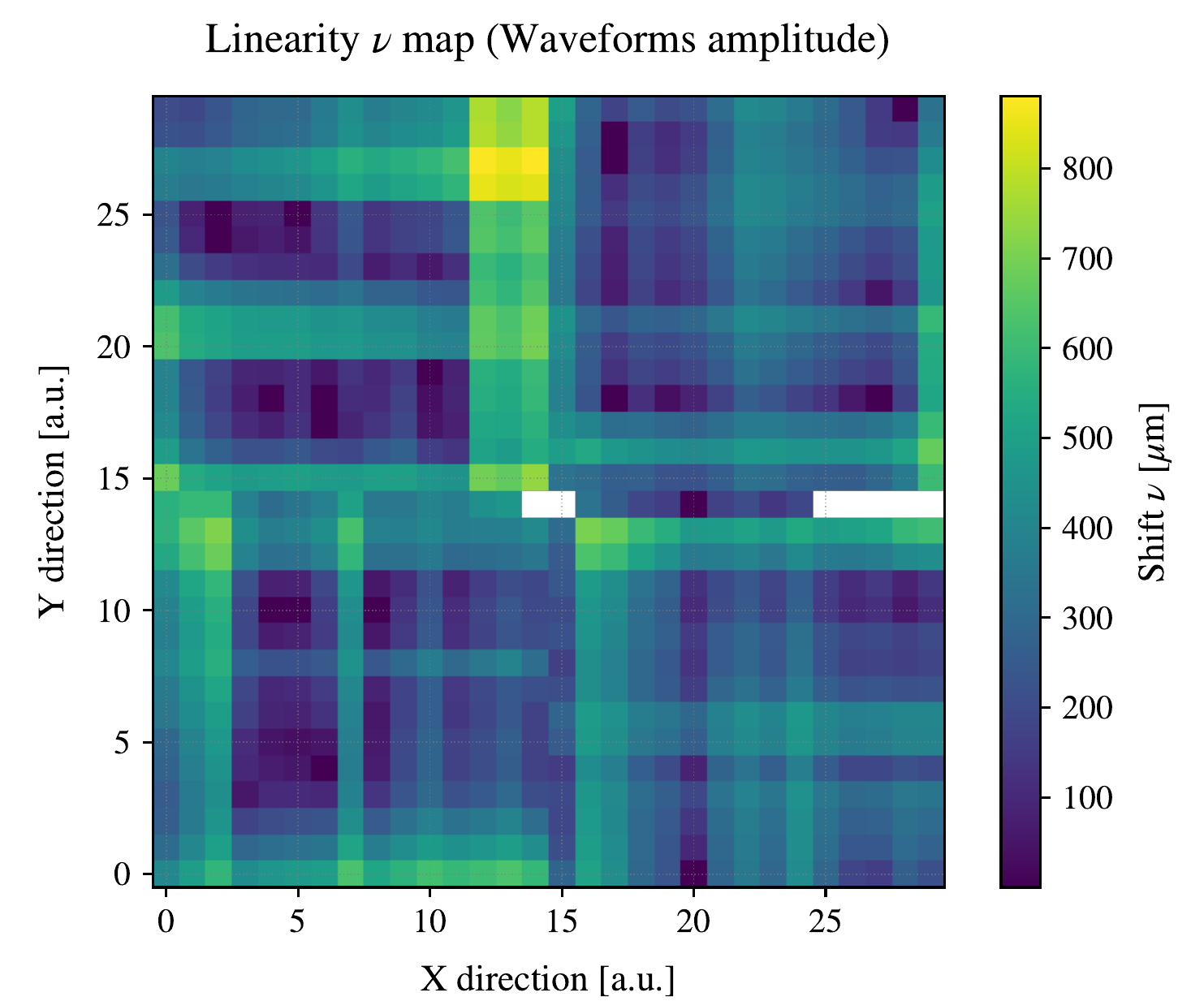}
    \caption{Linearity map of the SiPM surface scan using Rice fitting. Each bin represents a motor step.}
    \label{fig:lin_amp}
  \end{minipage}
\end{figure*}
The weighted averages for resolution and shift, computed from the amplitude reconstruction method, include only the central region of the camera, excluding the $1.5~\rm{mm}$ border where deformations prevent fit convergence. Results are summarized in Table~\ref{table:amp}.
\begin{table}[h]
\renewcommand{\arraystretch}{1.2}  
\centering
\begin{tabular}{p{0.4\linewidth}c}
\hline\hline
\textbf{Parameter}                 & \textbf{Average Rice fit result} \\ \hline
Average resolution $\bar{\sigma}$ & 81 $\pm$ 3 $\mu$m\\
Average shift $\bar{\nu}$ & 231 $\pm$ 4 $\mu$m\\
\hline
\end{tabular}
\caption{Average resolution and linearity parameter of 529 LED positions.}
\label{table:amp}
\end{table}
\par The reconstructed positions in Figure~\ref{fig:amplitude_method} clearly reveal the four SiPMs in the array. The gap between the upper SiPMs is visible, while the lower SiPMs appear less distinct due to varying lateral sizes in reconstruction (e.g., the bottom-right SiPM is reconstructed with a larger area). Additionally, a pincushion effect distorts the image towards the corners, as previously reported in~\cite{Acerbi2024}, likely due to high parasitic inductance and impedance mismatches.

\par The resolution map in Figure~\ref{fig:res_amp} shows degradation near the gaps between SiPM chips due to limited light collection (white pixels indicate insufficient photons for Rice fitting). Resolution also worsens towards the corners, influenced by the pincushion effect. Linearity, shown in Figure~\ref{fig:lin_amp}, exhibits strong deformations at the gap between the top SiPMs, with stripes crossing each SiPM corresponding to regions with larger $\nu$. 

\par Overall, the average resolution ($\sigma$) is $81 \pm 3~\rm{\mu m}$, confirming sub-millimetric precision for this PS-SiPM. Non-linearity effects, described by the shift $\nu$, indicate sensor accuracy at the level of a few hundred $\rm{\mu m}$.

\par These results are proper to the amplitude method reconstruction, through which waveform peaks are sampled and employed as inputs for equation~\ref{eq:LG-SiPM}. Even though other techniques were studied, such as extracting the charge by integrating the output waveforms for different time windows, we concluded that the amplitude-based approach yields the best average spatial resolution and linearity. 

\section{Conclusions}
This work introduces a statistical method to evaluate position-reconstruction performance for a 2x2 RGB-HD LG-SiPM array, assessing spatial resolution and linearity using iterative Rice fitting. The device achieves an average best resolution of $81\pm3~\rm{\mu m}$ ($190\pm7~\rm{\mu m}$ FWHM) and a few hundred $\mu m$ average shift for amplitude-based reconstruction (Table~\ref{table:amp}). The spatial resolution degrades near SiPM gaps and edges due to light collection limits and pincushion distortion. Results apply to $V_{ov}=5~\rm{V}$ at room temperature; lower temperatures or higher bias voltages improve resolution, while larger SiPM arrays worsen performance due to parasitics~\cite{Acerbi2024}. 

LG-SiPMs enable sub-millimeter CoG reconstruction, ideal for high-precision applications like Time Projection Chambers~\cite{Gola2020} and medical scanners. Their simple readout and reduced channel count support integration into systems requiring millimeter-scale resolution. The Rice-fitting technique can also assess other PS light sensors, given a nominal mapping between reconstructed positions and physical light source locations.

\section*{Aknowledgments}
This study was conducted as part of the POSiCS project, supported by the ATTRACT-Phase-2 initiative under the European Union’s H2020 program (Grant Agreement 101004462).  

The POSiCS project is run by a consortium between the Department of Physics (DPNC) and the PET Instrumentation and Neuroscience Lab (PINLab) of Universit\'e de Gen\`eve and the Fondazione Bruno Kessler (FBK).

The position-sensitive LG-SiPM is based on patented technology (EP3063559) developed by FBK.  
% \printbibliography

\bibliography{posics.bib}

\end{document}